\documentclass[11pt]{cernrep}

\newcommand{\sfrac}[2]{\mbox{\footnotesize $\frac{#1}{#2}$}} 

\usepackage{graphicx}
\usepackage{here}
\begin{document}
 \title{DSEs and pseudoscalar mesons: an aper\c{c}u}
\author{A.\,H\"oll,\footnotemark[2]\,\,\footnotemark[3] \ A.\,Krassnigg\footnotemark[2] \ and C.D.\,Roberts\footnotemark[2]\,\,\footnotemark[3]}
\institute{\footnotemark[2] \ Physics Division, Argonne National Laboratory, Argonne IL 60439, USA\\
\footnotemark[3] \ Fachbereich Physik, Universit\"at Rostock, D-18051 Rostock, Germany}
\maketitle
\begin{abstract}
An hallmark of present-day Dyson-Schwinger equation applications in had\-ron physics is the existence of a systematic and symmetry preserving truncation scheme.  This enables the proof of exact results; e.g., the leptonic decay constant of every pseudoscalar meson except the pion vanishes in the chiral limit.  Calculations using the scheme's leading-order truncation are reliable in the vector and flavour nonsinglet pseudoscalar channels.  In this rainbow-ladder truncation, an impulse approximation provides the consistent current for all six-point quark Schwinger functions.  That is well illustrated via the anomalous process $\pi^0 \to \gamma\gamma$.  Using two-, three- and four-point Schwinger functions calculated in rainbow-ladder truncation, the textbook value of the width is obtained algebraically and independent of model details if, and only if, the impulse approximation is used to describe the associated matrix element.  
\end{abstract}

\section{INTRODUCTION}
A number of brief accounts of the modern application of Dyson-Schwinger equations (DSEs) to hadron physics and QCD have recently appeared \cite{pcterice,kra,krb,racf,mrp}.  They have recapitulated on aspects of dynamical chiral symmetry breaking and confinement, and the expression of these phenomena in the hadron spectrum; and contain original contributions regarding: comparisons between DSE results and lattice simulations, and the mutual benefits of such comparisons \cite{pcterice,racf}; the pion and its excited states \cite{kra,krb}; and states with vacuum quantum numbers \cite{mrp}.  We have a modicum to add.

\section{GAP AND BETHE-SALPETER EQUATIONS}
We begin with QCD's renormalised gap equation
\begin{equation} 
\label{gendse} S(p)^{-1} = Z_2 \,(i\gamma\cdot p + m_{\rm bare}) +\, Z_1 
\int^\Lambda_q g^2 D_{\mu\nu}(p-q) \frac{\lambda^a}{2}\gamma_\mu S(q) 
\Gamma^a_\nu(q,p) \,,
\end{equation}
wherein: $D_{\mu\nu}(k)$ is the dressed-gluon propagator; $\Gamma^a_\nu(q,p)$
is the dressed-quark-gluon vertex; $m_{\rm bare}$ is the
$\Lambda$-de\-pen\-dent current-quark bare mass; and $\int^\Lambda_q :=
\int^\Lambda d^4 q/(2\pi)^4$ represents a translationally-invariant
regularisation of the integral, with $\Lambda$ the regularisation mass-scale.  We employ a Pauli-Villars scheme, which is implemented in Eq.\,(\ref{gendse}) by considering the quarks as minimally anticoupled ($g^{PV}=ig$) to additional massive gluons ($m_g^{PV}=\Lambda$).  This effects a tempering of the integrand that is expressed via a modification of the gluon propagator's ultraviolet behaviour:
\begin{equation}
\frac{1}{(p-q)^2} \to \frac{1}{(p-q)^2} - \frac{1}{(p-q)^2+\Lambda^2}\,,
\end{equation}
through which the integral's superficial linear divergence is regulated.  The regularising mass-scale is naturally removed to infinity as the last stage in the calculation of any renormalisable quantity.  In Eq.\,(\ref{gendse}), $Z_{1,2}(\zeta^2,\Lambda^2)$ are respectively the quark-gluon-vertex and quark wave function renormalisation constants, which depend on $\Lambda$ and the renormalisation point, $\zeta$, as does the mass renormalisation constant $Z_m(\zeta^2,\Lambda^2) = Z_4(\zeta^2,\Lambda^2)/Z_2(\zeta^2,\Lambda^2) $.  The gap equation's solution has the form
\begin{equation} 
 S(p)^{-1} =  i \gamma\cdot p \, A(p^2,\zeta^2) + B(p^2,\zeta^2) 
 \equiv \frac{1}{Z(p^2,\zeta^2)}\left[ i\gamma\cdot p + M(p^2)\right], 
\label{sinvp} 
\end{equation} 
where $M(\zeta^2) \equiv m(\zeta):= m_{\rm bare}(\Lambda)\, Z_m^{-1}(\zeta^2,\Lambda^2)$ is the running quark mass,\footnote{The dressed-quark mass function is renormalisation-point-independent; i.e., $M(p^2,\zeta^2)=M(p^2,\tilde\zeta^2)$.} and is obtained subject to the renormalisation condition 
\begin{equation} 
\label{renormS} \left.S(p)^{-1}\right|_{p^2=\zeta^2} = i\gamma\cdot p + 
m(\zeta)\,. 
\end{equation} 
The renormalisation scheme should be understood to yield mass-independent renormalisation constants.  
 
The gap equation illustrates the features and flaws of each DSE.  It is a
nonlinear integral equation for $S(p)$ and hence can yield much-needed
nonperturbative information.  However, the kernel involves the two-point
function $D_{\mu\nu}(k)$ and the three-point function
$\Gamma^a_\nu(q,p)$.  The gap equation is therefore coupled to the DSEs these
functions satisfy.  Those equations in turn involve higher $n$-point functions
and hence the DSEs are a tower of coupled integral equations with a tractable
problem obtained only once a truncation scheme is specified.  It is
unsurprising that the best known truncation scheme is the weak coupling
expansion, which reproduces every diagram in perturbation theory.  This
scheme is systematic and valuable in the analysis of large momentum transfer
phenomena because QCD is asymptotically free but it precludes any possibility
of obtaining nonperturbative information, which we identified as a key aspect
of the DSEs.

The gap equation describes the manner in which the propagation of a quark is affected by the medium being traversed, and the same is true of the gauge sector gap equations satisfied by gluon and ghost Schwinger functions.  The Bethe-Salpeter equations are the necessary complement as they provide the means by which one may understand how QCD's elementary excitations form the colour singlet bound states that are the observable manifestation of strong interaction dynamics.

An important example is the inhomogeneous Bethe-Salpeter equation for the axial-vector vertex:
\begin{equation}
\label{avbse}
\left[ \Gamma_{5\mu}^j(k;P) \right]_{tu} = Z_2 \left[\gamma_5 \gamma_\mu \frac{\tau^j}{2}\right]_{tu} + \int^\Lambda_q \left[ \chi_{5\mu}(q;P) \right]_{sr} \, K_{tu}^{rs}(q,k;P) \,,
 \end{equation}
where $\chi_{5\mu}(q;P)={\cal S}(q_1) \Gamma_{5\mu}(q;P){\cal S}(q_2)$ with ${\cal S}(\ell)={\rm diag}[S_u(\ell),S_d(\ell)]$ and, plainly, we focus on $SU(N_f=2)$ so that $\{\tau^j;j=1,2,3\}$ are Pauli matrices in flavour space.  The conventions of Eq.\,(\ref{avbse}) describe a vertex with total momentum $P$ and relative momentum $k$, whose outgoing quark and antiquark legs carry momenta $k_1=k+\eta P$ and $-k_2=-k + (1-\eta ) P$, respectively.  The appearance here of $\eta \in [0,1]$ expresses the arbitrariness in defining the relative momentum in a Poincar\'e covariant framework: physical observables will be independent of $\eta$ in a symmetry preserving truncation, as seen, e.g., in Ref.\,\cite{mr97}.  Of particular importance in Eq.\,(\ref{avbse}) is $K(q,k;P)$, the fully amputated dressed quark-antiquark scattering kernel.  This four-point Schwinger function also appears implicitly in QCD's gap equation because it is one piece of the kernel in the inhomogeneous integral equation satisfied by $\Gamma_\mu(q,p)$.

\section{PION AND RELATED MATTERS}
A primary measure of our comprehension of strong QCD is the possession of a veracious understanding of the pion, and for this a clear grasp of the nature of chiral symmetry and its dynamical breaking are vital.  In asymptotically free theories, chiral symmetry and its breaking are completely expressed via the axial-vector Ward-Takahashi identity:
\begin{equation} 
\label{avwtim} P_\mu \Gamma_{5\mu}^j(k;P)  = {\cal S}^{-1}(k_+)\, i 
\gamma_5\frac{\tau^j}{2} +  i \gamma_5\frac{\tau^j}{2}\, {\cal S}^{-1}(k_-) - i 
{\cal M}(\zeta) \,\Gamma_5^j(k;P) - \Gamma_5^j(k;P)\,i {\cal M}(\zeta), 
\end{equation} 
wherein the new elements are the current-quark mass matrix: ${\cal M}(\zeta) = 
{\rm diag}[m_u(\zeta),m_d(\zeta)]$, and the pseudoscalar vertex
\begin{equation}
\label{psbse}
\left[ \Gamma_{5}^j(k;P) \right]_{tu} = Z_4 \left[\gamma_5 \frac{\tau^j}{2}\right]_{tu} + \int^\Lambda_q \left[ \chi_{5}(q;P) \right]_{sr} \, K_{tu}^{rs}(q,k;P) \,.
\end{equation}

The last two terms of Eq.\,(\ref{avwtim}) vanish in the chiral limit.  The identity is then a concise statement of chiral symmetry, a consequence of which is plainly an intimate relation between the three-point function in Eq.\,(\ref{avbse}) and the two-point function in Eq.\,(\ref{gendse}).  Since dynamical chiral symmetry breaking is an essentially nonperturbative phenomenon, satisfying our measure of comprehension requires that there exist a nonperturbative and systematic truncation scheme which preserves identities such as Eq.\,(\ref{avwtim}). 

\subsection{Model-independent results}
At least one such scheme exists \cite{bender,detmold}: a dressed-loop expansion of the dressed-quark-gluon vertices that appear in the half-amputated dressed-quark-anti\-quark scattering kernel: $(SS) K$, and that is sufficient to arrive via the DSEs at exact results in QCD.  For example, it is a general property that the axial-vector vertex, obtained via Eq.\,(\ref{avbse}), exhibits a pole whenever $P^2= - m^2_{\pi_n}$, where $m_{\pi_n}$ is the mass of any pseudoscalar $u$-$d$ meson; viz., 
\begin{equation} 
\left. \Gamma_{5 \mu}^j(k;P)\right|_{P^2+m_{\pi_n}^2 \approx 0} = 
\mbox{regular\ terms} +  \frac{f_{\pi_n} \, P_\mu}{P^2 + 
m_{\pi_n}^2} \Gamma_{\pi_n}^j(k;P)\,, \label{genavv} 
\end{equation} 
with $\Gamma_{\pi_n}^j(k;P)$ being the $0^{-+}$ bound state's Bethe-Salpeter 
amplitude: 
\begin{equation} 
 \Gamma_{\pi_n}^j(k;P)= \tau^j \gamma_5 \left[ i E_{\pi_n}(k;P) 
+ \gamma\cdot P F_{\pi_n}(k;P)  +\,  \gamma\cdot k \,k \cdot P\, G_{\pi_n}(k;P) + 
\sigma_{\mu\nu}\,k_\mu P_\nu \,H_{\pi_n}(k;P)  \right], \label{genpibsa} 
\end{equation} 
and $f_{\pi_n}$, its leptonic decay constant
\begin{equation} 
\label{fpin} f_{\pi_n} \,\delta^{ij} \,  P_\mu = Z_2\,{\rm tr} \int^\Lambda_q 
\sfrac{1}{2} \tau^i \gamma_5\gamma_\mu\, {\cal S}(q_+) \Gamma^j_{\pi_n}(q;P) 
{\cal S}(q_-) \,, 
\end{equation} 
where the trace is over colour, flavour and spinor indices.  Equation 
(\ref{fpin}) is the precise expression in quantum field theory for the 
pseudo\textit{vector} projection of the meson's Bethe-Salpeter wave function 
onto the origin in configuration space.  The appearance of $Z_2$ guarantees that $f_{\pi_n}$ is gauge invariant, and independent of the regularisation scale and renormalisation point.

The lowest mass pole contribution to Eq.\,(\ref{genavv}), denoted by $n=0$, is the ground state pion, which receives the spectroscopic assignment $N\, ^{2 s +1}\!L_J = 1 \, ^1\!S_0$ in the naive $q \bar q$ quark model.  The $n\geq 1$ pseudoscalar meson poles correspond to those excited states of the pion that would receive the 
spectroscopic assignments $(n+1)\, ^1\!S_0$ in the quark model. 
 
The pseudoscalar vertex, $\Gamma_5^j(k;P)$, which appears in 
Eq.\,(\ref{avwtim}), also exhibits such a pole: 
\begin{equation} 
 \left. i \Gamma_{5}^j(k;P)\right|_{P^2+m_{\pi_n}^2 \approx 0} = 
\mbox{regular\ terms} + \frac{ \rho_{\pi_n} }{P^2 + m_{\pi_n}^2}\, 
\Gamma_{\pi_n}^j(k;P)\,,\label{genpvv} 
\end{equation} 
and here the residue
\begin{equation} 
\label{cpres} i  \rho_{\pi_n}\!(\zeta)\, \delta^{ij}  = Z_4\,{\rm tr} 
\int^\Lambda_q \sfrac{1}{2} \tau^i \gamma_5 \, {\cal S}(q_+) 
\Gamma^j_{\pi_n}(q;P) {\cal S}(q_-) 
\end{equation} 
expresses the pseudo\textit{scalar} projection of the meson's Bethe-Salpeter wave function onto the origin in configuration space.  The appearance of $Z_4$ guarantees that $\rho_{\pi_n}$ is gauge invariant, independent of the regularisation scale, and evolves correctly under the renormalisation group.

In QCD it therefore follows from Eqs.\,(\ref{genavv}) and (\ref{genpvv}), as a necessary consequence of chiral symmetry and its dynamical breaking, that for any pseudoscalar $u$-$d$ meson 
\cite{kra,krb,mrt98} 
\begin{equation} 
\label{gmorgen} f_{\pi_n} m_{\pi_n}^2 = [ m_u(\zeta) + m_d(\zeta) ] \, 
\rho_{\pi_n}(\zeta)\,. 
\end{equation} 
(This expression is valid independent of the magnitude of the current-quark mass and, following Ref.\,\cite{mr97}, the generalisation to $SU(N_f)$-flavour is straightforward.)  The so-called Gell-Mann--Oakes--Renner relation for the 
ground state pion appears as a corollary of Eq.\,(\ref{gmorgen}) 
\cite{mrt98,langfeld}, and another important corollary of Eq.\,(\ref{gmorgen}), valid 
for pseudoscalar mesons containing at least one heavy-quark, is described in 
Refs.\,\cite{hqlimit}.  In addition, the proof of Eq.\,(\ref{gmorgen}) establishes \cite{kra} that in the chiral limit 
\begin{equation} 
\label{fpiniszero} 
f^0_{\pi_{n\neq 0}}:= \lim_{m \to 0} \, f_{\pi_{n\neq 0}} = 0 \,; 
\end{equation} 
namely, in the chiral limit the leptonic decay constant vanishes for \textit{every} 
pseudoscalar meson \textit{except} the ground state pion.  This result in particular is a constraint on models and theoretical methods used to search for exotics and hybrids. 

\subsection{\textit{Ab initio} calculations}
The leading-order term in the scheme of Refs.\,\cite{bender,detmold} is the renormalisation-group-improved rainbow-ladder truncation, which is expressed in the Bethe-Salpeter equations via
\begin{equation} 
\label{ladder} 
K_{tu}^{rs}(q,k;P) = - 4 \pi \alpha(Q^2) D_{\rho\sigma}^{\rm free}(Q)\, 
\left[\rule{0mm}{0.7\baselineskip} 
        \frac{\lambda^a}{2}\gamma_\rho \right]_{ts} 
\left[\rule{0mm}{0.7\baselineskip} 
        \frac{\lambda^a}{2}\gamma_\sigma\right]_{r u}\!,
\end{equation}
with $\alpha(Q^2)$ an \textit{operative} coupling, and in the truncated gap equation through:
\begin{equation} 
\label{rainbowDSE}
S(p)^{-1} = Z_2 \,(i\gamma\cdot p + m_{\rm bare}) +\, \int^\Lambda_q \, 
4\pi\alpha((p-q)^2)\,D^{\rm free}_{\mu\nu}(p-q) \frac{\lambda^a}{2}\gamma_\mu 
S(q) \frac{\lambda^a}{2}\gamma_\nu \,.
\end{equation}  

This truncation can be used to correlate lattice data on dressed-quark and -gluon Schwinger functions \cite{mandar}.  It is also a reliable tool in the study of vector and flavour nonsinglet pseudoscalar channels \cite{detmold}, and has been refined and exploited by Maris and Tandy in a series of articles\footnote{An exemplary success was their prediction \protect\cite{maristandypion} of the electromagnetic pion form factor \protect\cite{volmer}.  In addition, see Refs.\,\protect\cite{maristandyO,maristandy} and references therein, and the review of their substantial contributions in 
Ref.\,\protect\cite{revpieter}.}  via a one-parameter model for the operative coupling: 
\begin{equation}
\label{alphamt} 
\frac{\alpha(Q^2)}{Q^2} = \frac{\pi}{\omega^6}\, D\, Q^2 \, {\rm e}^{-Q^2/\omega^2} 
+ \,\frac{ \pi\, \gamma_m } { \frac{1}{2}\ln\left[\tau + \left(1 + 
Q^2/\Lambda_{\rm QCD}^2\right)^2\right]} \, {\cal F}(Q^2) \,, \label{gk2} 
\end{equation} 
wherein ${\cal F}(Q^2)= [1 - \exp(-Q^2/[4 m_t^2])]/Q^2$, $m_t$ $=$ 
$0.5\,$GeV; $\tau={\rm e}^2-1$; $\gamma_m = 12/25$; and $\Lambda_{\rm QCD} 
=0.234\,$GeV.  This form expresses the interaction as a sum: the 
second term ensures that perturbative behaviour is preserved at short-range; 
and the first makes provision for enhancement (strong coupling) at long-range.  The true parameters in Eq.\,(\ref{gk2}) are $D$ and $\omega$, which together determine 
the integrated infrared strength of the rainbow-ladder kernel.  However, they 
are not independent: in fitting to a selection of observables, a change in 
one is compensated by altering the other; e.g., on the domain 
$\omega\in[0.3,0.5]\,$GeV, the fitted observables are approximately constant 
along the trajectory $\omega \,D = (0.72\,{\rm GeV})^3$ \cite{raya}. Hence 
Eq.\,(\ref{gk2}) is a one-parameter model.  This correlation: a reduction in 
$D$ compensating an increase in $\omega$, ensures a fixed value of the interaction's
integrated infrared strength.  It also defines a single dressed-glue mass-scale that characterises infrared gluodynamics:
\begin{equation}
m_g = 720\,{\rm MeV}\,.
\end{equation}

Using Maris and Tandy's model for the operative coupling, the rainbow-ladder truncation yields \cite{maristandyO} $ f_{\pi_0} = 0.092$\,{\rm GeV}; $m_{\pi_0} = 0.14$\,{\rm GeV}; $\rho_{\pi_0} = (0.81\,{\rm GeV})^2$, at a current-quark mass $m_d(1\,{\rm GeV})= m_u(1\,{\rm GeV})= 
5.5\,$MeV.  This model is currently being employed in a rainbow-ladder analysis of $n\neq 0$ pseudoscalar mesons, and it has been reported \cite{kra,krb} that when only the $E_{\pi_1}(k;P)$ term in Eq.\,(\ref{genpibsa}) is retained,\footnote{With this amplitude alone one obtains values for $m_{\pi_0}$, $f_{\pi_0}$ that are approximately $25$\% smaller than experiment.} one obtains: $m_{\pi_1} = 1.1\,{\rm GeV}$ cf.\ $m_{\pi_1}^{\rm expt.} = 1.3\pm 0.1 \,{\rm GeV}$; and $f_{\pi_{n=1}} \leq 1.5\,{\rm MeV}$ at the physical (nonzero) current-quark mass.  The latter result indicates that Eq.\,(\ref{fpiniszero}) is a good approximation even at real-world current-quark masses and thereby emphasises its usefulness as a constraint.
 
\section{ABELIAN ANOMALY}
A correct description of the neutral pion's two-photon decay is another stringent test of one's understanding of strong QCD, and that understanding again relies fundamentally on a true comprehension of chiral symmetry and its dynamical breakdown.  To illustrate this we consider the renormalised impulse approximation to the axial-vector--vector--vector (AVV) vertex:
\begin{equation}
\label{Tmnr}
T^3_{\mu\nu\rho}(k_1,k_2) = {\rm tr}\int_\ell^M {\cal S}(\ell_{0+}) \, \Gamma^3_{5\rho}(\ell_{0+},\ell_{-0}) \, {\cal S}(\ell_{-0}) \, i{\cal Q}\Gamma_\mu(\ell_{-0},\ell) \, {\cal S}(\ell) \, i {\cal Q}\Gamma_\nu(\ell,\ell_{0+})\,,
\end{equation}
where $\ell_{\alpha\beta}=\ell+\alpha k_1+\beta k_2$ and the electric charge matrix ${\cal Q}={\rm diag}[e_u,e_d]=e\,{\rm diag}[2/3,-1/3]$.  In Eq.\,(\ref{Tmnr}), the dressed-quark propagator is determined by the rainbow gap equation, and the dressed vector and axial-vector vertices are obtained from the dressed-ladder truncation of the relevant inhomogeneous Bethe-Salpeter equations.  It is apparent from Eq.\,(\ref{genavv}) that for $(k_1+k_2)^2+m_\pi^2\approx 0$; namely, in the neighbourhood of the pion pole, $T^3_{\mu\nu\rho}(k_1,k_2)$ is dominated by the $\pi^0 \to \gamma \gamma$ coupling and therefore plays a role in describing that decay.

The bare AVV vertex exhibits a superficial linear divergence and as with all other Schwinger functions it must be rigorously defined via a translationally invariant regularisation scheme.  In this case the appropriate Pauli-Villars prescription corresponds to minimally anticoupling the photon to additional flavoured quarks with a large mass $m_q^{PV}=M$.  To elucidate we introduce
\begin{equation}
\tilde T^3_{\mu\nu\rho}(k_1,k_2;\hat m) := {\rm tr}\int_\ell {\cal S}_{\hat m}(\ell_{0+}) \, \Gamma^{3\,\hat m}_{5\rho}(\ell_{0+},\ell_{-0}) \, {\cal S}_{\hat m}(\ell_{-0}) \, i {\cal Q} \Gamma^{\hat m}_\mu(\ell_{-0},\ell) \, {\cal S}_{\hat m}(\ell) \, i {\cal Q}\Gamma^{\hat m}_\nu(\ell,\ell_{0+})\,,
\end{equation}
where $\hat m$ is the renormalisation point invariant current-quark mass and every element in the integrand is calculated using this current-quark mass in the appropriate rainbow-ladder truncated DSE.  In this case Eq.\,(\ref{Tmnr}) is explicitly
\begin{equation}
T^3_{\mu\nu\rho}(k_1,k_2;\hat m) = \tilde T^3_{\mu\nu\rho}(k_1,k_2;\hat m) - \tilde T^3_{\mu\nu\rho}(k_1,k_2;M)\,,
\end{equation}
with $M \to \infty$ as the last step in the calculation.

\begin{figure}[t]
\begin{center}
\includegraphics[width=0.8\textwidth]{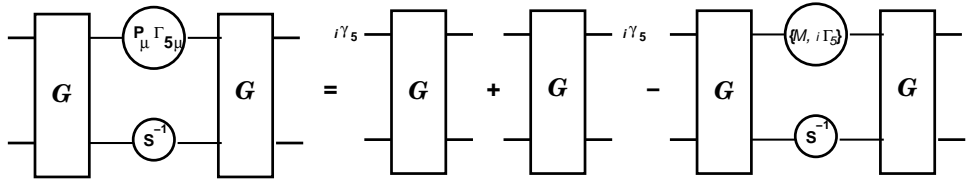} 
\caption{\label{figAVWTI} This axial-vector Ward-Takahashi identity is an extension of Eq.\,(\protect\ref{avwti0}).  It is valid if, and only if: the dressed-quark propagator, $S$, is obtained from Eq.\,(\ref{rainbowDSE}); the axial-vector vertex, $\Gamma_{5\mu}$, is obtained from Eq.\,(\protect\ref{avbse}) with the kernel constructed from $S$ and Eq.\,(\protect\ref{ladder}); the pseudoscalar vertex is constructed analogously; and the unamputated renormalised quark-antiquark scattering matrix: $G= (SS) + (SS)K(SS) + (SS)K(SS)K(SS)+ [\ldots]$, is constructed from the elements just described.}
\end{center}
\end{figure}

We now focus on the chiral limit, $\hat m = 0$, and consider the divergence of the AVV vertex at $P^2= (k_1+k_2)^2 \approx 0$; viz., in the neighbourhood of the massless pion's pole:
\begin{eqnarray}
\lefteqn{
\left. P_\rho\,T^3_{\mu\nu\rho}(k_1,k_2;0)\right|_{P^2= (k_1+k_2)^2 \approx 0} = 
P_\rho \tilde T^3_{\mu\nu\rho}(k_1,k_2;\hat m) - P_\rho \tilde T^3_{\mu\nu\rho}(k_1,k_2;M)}\\
\nonumber
& = & Z_2(\zeta^2,M^2) \, {\rm tr}\int_\ell \gamma_5 \gamma_\nu {\cal Q} \frac{\tau^3}{2}\left\{ \,{\cal S}_{\hat m}(\ell_{-0}) \, i{\cal Q}\Gamma^{\hat m}_\mu(\ell_{-0},\ell) \, {\cal S}_{\hat m}(\ell) - {\cal S}_{M}(\ell_{-0}) \, i{\cal Q}\Gamma^{M}_\mu(\ell_{-0},\ell) \, {\cal S}_{M}(\ell)\right\}\\
\nonumber
& -& Z_2(\zeta^2,M^2) \, {\rm tr}\int_\ell \gamma_5 \gamma_\mu {\cal Q} \frac{\tau^3}{2}\left\{ \,{\cal S}_{\hat m}(\ell) \, i{\cal Q}\Gamma^{\hat m}_\nu(\ell,\ell_{0+}) \, {\cal S}_{\hat m}(\ell_{0+}) - {\cal S}_{M}(\ell) \, i{\cal Q}\Gamma^{M}_\nu(\ell,\ell_{0+}) \, {\cal S}_{M}(\ell_{0+})\right\}\\
&+ & \, 2 M(\zeta) \, {\rm tr}\int_\ell {\cal S}_{M}(\ell_{0+}) \, i\Gamma^{3\,M}_{5}(\ell_{0+},\ell_{-0}) \, {\cal S}_{M}(\ell_{-0}) \, i{\cal Q}\Gamma^{M}_\mu(\ell_{-0},\ell) \, {\cal S}_{M}(\ell) \, i {\cal Q}\Gamma^{M}_\nu(\ell,\ell_{0+})\,.
\label{PTmnr}
\end{eqnarray}
Equation\,(\ref{PTmnr}) follows from the vector vertex analogue of Eq.\,(\ref{avbse}) and the axial-vector Ward-Takahashi identity depicted in Fig.\,\ref{figAVWTI},\footnote{The relevance of this Ward-Takahashi identity is plain if one re-expresses the vector vertices through: $\Gamma_\mu  = Z_2 \gamma_\mu \, G\, (SS)^{-1}$, where $G$ is the unamputated renormalised rainbow-ladder-dressed quark-antiquark scattering matrix.} which is an extension of
\begin{equation}
\label{avwti0}
P_\mu {\cal S}(k_+) \,\Gamma_{5\mu}^j(k;P)\, {\cal S}(k_-)  =  i 
\gamma_5\frac{\tau^j}{2} \, {\cal S}(k_-) +  {\cal S}(k_+)\, i \gamma_5\frac{\tau^j}{2} - {\cal S}(k_+) \{{\cal M}(\zeta)\, , \,i\Gamma_5^j(k;P)\} {\cal S}(k_-)\,,
\end{equation}
and can be derived following the method in Ref.\,\cite{bicudo}: the identity is valid if, and only if, every dressed-quark propagator that appears therein is obtained from the rainbow DSE, Eq.\,(\ref{rainbowDSE}), and the accompanying dressed vertices are determined from the consistent ladder Bethe-Salpeter equation; i.e., the equation constructed using Eq.\,(\ref{ladder}) and the rainbow quark propagator.   It is plain that in the chiral limit Eq.\,(\ref{PTmnr}) does not contain terms which depend on the light-quark mass.

Consider the first term on the r.h.s.\ of Eq.\,(\ref{PTmnr}).  The complete polarisation integral is convergent, by virtue of our regularisation, and must yield a rank-two Lorentz pseudotensor depending on only one four-vector argument, $k_1$.  A nonzero value is impossible.  The second term vanishes for the same reason.  

Let's turn to the last term.  The only relevant contribution appears in the limit $M\to \infty$, in which case all dressing is suppressed and we are left to evaluate
\begin{eqnarray}
\nonumber
\lefteqn{
T_{\mu\nu}(k_1,k_2;M)}\\
& =& 2 M \, {\rm tr}\int\! \frac{d^4\ell}{(2\pi)^4}\, \frac{1}{i\gamma\cdot \ell_{0+} + M}\, i\gamma_5\frac{\tau^3}{2}\, \frac{1}{i\gamma\cdot\ell_{-0}+ M}\, i{\cal Q}\gamma_\mu \, \frac{1}{i\gamma\cdot \ell + M} \, i {\cal Q}\gamma_\nu\\
& = & N_c\, (e_u^2-e_d^2) \, M \, {\rm tr}_D\int\!\frac{d^4\ell}{(2\pi)^4}\, \frac{1}{i\gamma\cdot \ell_{0+} + M}\, i\gamma_5\, \frac{1}{i\gamma\cdot\ell_{-0}+ M}\, i\gamma_\mu \, \frac{1}{i\gamma\cdot \ell + M} \, i \gamma_\nu\\
& = & 4 \, i e^2 \,  \varepsilon_{\mu\nu \rho\sigma}\, k_{1\rho}\, k_{2\sigma}\,
\int\!\frac{d^4\ell}{(2\pi)^4}\,\frac{M^2}{(\ell_{0+}^2 + M^2) \, (\ell_{-0}^2+M^2)\, (\ell^2 +M^2)}\\
&=& i \frac{\alpha}{\pi} \, \varepsilon_{\mu\nu \rho\sigma}\, k_{1\rho}\, k_{2\sigma}\,
\int_0^1\!dx \int_0^x\! dy\, \frac{M^2}{M^2+  x y \,(k_1+ k_2)^2}\,.
\end{eqnarray} 

Recall now that at any order in the nonperturbative, systematic, symmetry-preserving DSE truncation scheme of Refs.\,\cite{bender,detmold}, dynamical chiral symmetry breaking entails that in the chiral limit the divergence of the axial-vector vertex is dominated by the pion pole in the neighbourhood of $P^2=0$. This result is expressed in Eq.\,(\ref{genavv}).  It follows that in the rainbow-ladder truncation, which is the leading-order in that scheme, 
\begin{eqnarray}
\nonumber \lefteqn{
f_\pi \, T_{\mu\nu}(k_1,k_2):= \left. P_\rho\,T^3_{\mu\nu\rho}(k_1,k_2)\right|_{P^2=(k_1+k_2)^2 \approx 0} }\\
& = & f_\pi {\rm tr}\int_\ell^M {\cal S}(\ell_{0+}) \, \Gamma_{\pi^0}(\ell_{0+},\ell_{-0}) \, {\cal S}(\ell_{-0}) \, i{\cal Q}\Gamma_\mu(\ell_{-0},\ell) \, {\cal S}(\ell) \, i {\cal Q}\Gamma_\nu(\ell,\ell_{0+})\,. \label{pi0gg}
\end{eqnarray}
Consequently, in the chiral limit, the Bose-symmetrised amplitude for $\pi^0 \to \gamma\gamma$ decay 
\begin{equation}
\label{Apigg}
{\cal A}^{\pi^0 \to \gamma\gamma}_{\mu\nu} = T_{\mu\nu}(k_1,k_2) + T_{\nu\mu}(k_2,k_1) = i \frac{\alpha}{\pi f_\pi}\, \varepsilon_{\mu\nu \rho\sigma}\, k_{1\rho}\, k_{2\sigma}\,.
\end{equation}

In Eq.\,(\ref{Apigg}) we have arrived at the textbook result with no dependence on model details; viz., independent of the detailed form of the operative coupling, Eq.\,(\ref{alphamt}), and thereby provided an explanation of the numerical analysis reported in Ref.\,\cite{pmpi0}.  In doing so we have made plain that the impulse approximation expressed in Eqs.\,(\ref{Tmnr}) and (\ref{pi0gg}) is the precise and only truncation of the six-point Schwinger functions relevant to this decay which is consistent with the rainbow-ladder truncation.  It follows that the ``seagull terms'' identified in Ref.\,\cite{banff} are absent at this order but may appear in a consistent treatment of the process at an higher order of truncation.  Moreover, it is generally true that seagull terms are absent in the consistent rainbow-ladder analysis of any six-point quark Schwinger function.

\section{EPILOGUE}
There are many additional applications of interest to this audience, of which a variety are reviewed in Refs.\,\cite{revpieter,revbasti}, and one may say that the 
keystones of success in modern DSE usage are: an appreciation and expression of the momentum-dependence of dressed-parton propagators at infrared length-scales; and the identification and employment of a nonperturbative, systematic, symmetry preserving truncation scheme. 

One topic of material importance, worth mentioning explicitly here, is the pion's valence quark distribution.  The DSE calculation \cite{cdruvx} agrees with  perturbative QCD's prediction \cite{Brodskyuvx}; i.e., $u_v^\pi(x) \propto (1-x)^2$ in the valence region,\footnote{In connection with calculations of $u_v^\pi(x)$ it must be borne in mind that the pion's Bethe-Salpeter amplitude is momentum dependent at every regularisation scale.  This is an essential feature of QCD and any calculation which overlooks it is immaterial.} but this is inconsistent with extant data \cite{piNDY} and a recent transverse lattice calculation \cite{dalley}.  The discrepancy between data and the expectation from perturbative QCD cannot be resolved  using contemporary simulations of lattice-regularised QCD \cite{wally}.  It raises difficult questions and has prompted proposals for remeasurement of the distribution function \cite{roy}.  

We have omitted a discussion of contemporary challenges confronting the application of DSEs in hadron physics and QCD.  Important amongst these are: the development of a veracious picture of the infrared behaviour of the Bethe-Salpeter kernel; and the extension of the framework to processes involving baryons.  This list is not exhaustive.  Nonetheless, comments on the former may be found in Refs.\,\cite{racf,mandar,hawes,alekseev,blochmr} and the latter in Refs.\,\cite{blochbaryon}.  

\vskip1cm
\noindent

\section*{ACKNOWLEDGMENTS}

CDR warmly thanks the organisers of this Light Cone Workshop: ``Hadrons and Beyond,'' at the UK Institute for Particle Physics Phenomenology, for their hospitality and support, and particularly Simon Dalley and Linda Wilkinson.
This work was also supported by: 
{\it FWF Erwin-Schr\"odinger-Auslandsstipendium Nr.}\ J2233-N08;
Department of Energy Nuclear Physics Division contract no.\ W-31-109-ENG-38;
National Science Foundation contract no.\ INT-0129236; 
and the A.\,v.\ Humboldt Foundation via a F.\,W.\ Bessel Research Award.

\end{document}